\documentclass[conference]{IEEEtran}
\IEEEoverridecommandlockouts
\pdfoutput=1

\usepackage{cite}
\usepackage{amsmath,amssymb,amsfonts}

\usepackage[noend]{algpseudocode}
\usepackage{algorithmicx,algorithm}

\usepackage{graphicx}
\usepackage{textcomp}
\usepackage{xcolor}

\usepackage{subfigure}
\usepackage[hyphens]{url}
\usepackage[hyphenbreaks]{breakurl}
\usepackage{array}
\usepackage{multirow}
\usepackage{balance}

\def\BibTeX{{\rm B\kern-.05em{\sc i\kern-.025em b}\kern-.08em
    T\kern-.1667em\lower.7ex\hbox{E}\kern-.125emX}}
\begin{document}

\title{Monitoring and Improving Personalized Sleep Quality from Long-Term Lifelogs}

\author{\IEEEauthorblockN{Wenbin Gan, Minh-Son Dao, Koji Zettsu}
\IEEEauthorblockA{\textit{Big Data Integration Research Center} \\
\textit{National Institute of Information and Communications Technology (NICT)}\\
Tokyo, Japan \\
$\{wenbingan,dao,zettsu\}@nict.go.jp$}
}

\maketitle

\begin{abstract}
Sleep plays a vital role in our physical, cognitive, and
psychological well-being. Despite its importance, long-term monitoring of personalized sleep quality (SQ) in real-world contexts is still challenging. Many sleep researches are still developing clinically and far from accessible to the general public. Fortunately,  wearables and IoT devices provide the potential to explore the sleep insights from multimodal data, and have been used in some SQ researches. However, most of these studies analyze the sleep related data and present the results in a delayed manner (i.e., today's SQ obtained from last night's data), it is sill difficult for individuals to know how their sleep will be before they go to bed and how they can proactively improve it. To this end, this paper proposes a computational framework to monitor the individual SQ based on both the objective and subjective data from multiple sources, and moves a step
further towards providing the personalized feedback to improve the
SQ in a data-driven manner. The feedback is implemented by referring the insights from the PMData dataset based on the  discovered patterns between life events and different
levels of SQ. The deep learning based personal SQ model (PerSQ), using the long-term heterogeneous data and considering the carry-over effect, achieves higher prediction performance compared with baseline models. A case study also shows reasonable results for an individual to monitor and improve the SQ in the future.

\end{abstract}


\section{Introduction}
\begin{figure}[!tbp]
\centerline{\includegraphics[width=.4\textwidth]{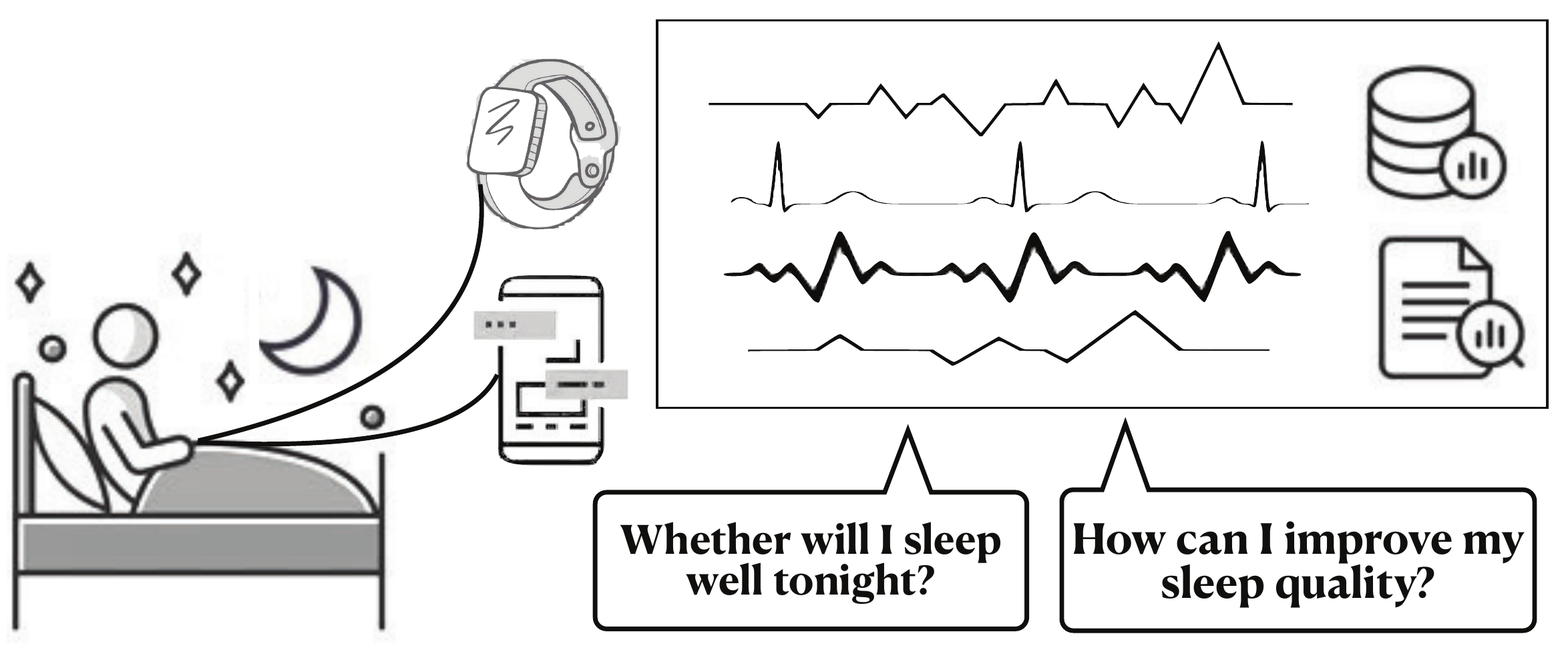}}
\caption{Research questions in this paper.}
\label{Fig:question}
\end{figure}

Sleep is of great importance for us human beings to maintain health. This complex physiological process is a regular part of our daily routine, allows our body and mind to recharge, become refreshed and alert 
\cite{mendoncca2019review}. A good night's sleep is closely linked to a greater level of physical, cognitive, and psychological well-being 
\cite{medic2017short,crivello2019meaning}. 
While getting enough sleep is definitely important, good SQ is also essential for a good night’s sleep. 
In this paper we focus on monitoring the long-term sleep qualities of individuals using daily logs and providing the personalized lifestyle feedback to improve the sleep qualities (Research questions are shown in Figure \ref{Fig:question}).

Sleep covers nearly one third of the individual lifespan; given its significant role for human lives and its high impact on our health,  
a large amount of work has been done to explore the quantitative sleep measures \cite{crivello2019meaning,mendoncca2019review}, the modeling of factors influencing sleep, such as physical activities \cite{sathyanarayana2016impact}, serious diseases \cite{medic2017short} and lifestyle \cite{shochat2012impact}. However, many sleep researches are still developing clinically and far from accessible to the general public \cite{upadhyay2020personalized}. For most people, it may be possible to simply self-assess the last night's sleep by observing the mental status and physical conditions after getting up, but it is still difficult to know how their sleep will be before they go to bed and how they can improve it \cite{pandey2020personalized,upadhyay2020personalized}.

Some clinical tools for sleep monitoring, such as the Polysomnography (PSG), can provide a quantitative insight of sleep and diagnose sleep disorders \cite{chang2020isleep}. PSG-based sleep studies generally perform overnight while being continually observed by a qualified technician using various medical sensors. Although such studies archive quite precise results for SQ measurement, they are expensive and unusual for an average person and thus only limited to clinical settings. As an inexpensive  alternation, actigraphy has been used as a popular non-invasive method to monitor human sleep; this method uses wristwatch-like wearables to continuously record human body movements overnight and assess the SQ. 
With the increasingly popularity of wearables and IoT devices, the multimodal data streams and daily events can be captured using these devices \cite{gan2022iot}, providing a premising way to jointly analyzing heterogeneous data sources to yield more accurate monitoring of individual SQ from a computational perspective \cite{pandey2020personalized,park2019learning}. 

Existing work on SQ measurement is  
mainly conducted at long-term and per-night level in terms of the duration of the measurements lasts \cite{bai2012will}. The former usually conducts the measurements over a relatively long time interval based on some self-rated questionnaires, such as the Pittsburgh SQ Index (PSQI) \cite{buysse1989pittsburgh}. The later performs the per-night SQ prediction on a daily basis \cite{bai2012will}. Compared with the long-term methods, the per-night methods provide the severity of a particular sleep problem more timely. 
Despite this benefit, most of the per-night SQ measurement methods do not consider the inter-dependencies between the coming tonight's sleep and the sleep conditions in the previous days (the carry-over effect), which may potentially cause the performance degradation. 
Besides, most of these two categories of studies analyze the sleep related data and present the results in a delayed
manner (i.e., today’s SQ obtained from last night’s data); it will be more useful to inform users their potential SQ before they go to bed and give feedback on action taking to improve it.      
Moreover, the majority of existing researches
build the population-based models 
to score user sleep without giving them the personalized feedback for improving their future SQ. Considering that individuals have varying requirements and reactions to sleep, it is necessary to monitor the individual SQ and provide the personalized lifestyle feedback to improve the SQ. 

To this end, this paper proposes a holistic framework to monitor the individual SQ based on both the objective and subjective lifelog data from multiple sources, and moves a step further to provide the personalized feedback to improve the SQ in a data-driven manner. It infers the individual SQ very timely before the bedtime using the long-term data to model the carry-over effect. Moreover, it builds the connections between SQ and the everyday life events, and actively feedbacks the individuals to bring the potential SQ improvement based on these connections.    
The contributions of this work are threefold:
\begin{itemize}
\item We propose an computational framework for monitoring the individual SQ and going a step further to provide the personalized feedback to improve the SQ.
\item We present the PerSQ model with high performance to monitor the individual SQ timely using the long-term data to model the carry-over effect.
\item We discover the insights of patterns between life events and different
levels of SQ, and use them to feedback the individuals to bring the potential SQ improvement. 
\end{itemize}


\section{Related Work}

\subsection{SQ Monitoring}
Sleep is a complex physiological process \cite{mendoncca2019review}. Although  there is still a lack of an established definition for the term SQ, researchers have explored to measure SQ from both clinical and technical fields using various quantitative metrics, such as 
total sleep time, sleep onset latency, degree of fragmentation, total wake time, and sleep efficiency \cite{crivello2019meaning}. It is well acknowledged that accurate SQ monitoring is essential for the early detection of sleep deprivation and insomnia. 

In general, SQ studies are mainly conducted from both objective and subjective perspectives in terms of the data source for the measurement. 
Sleep diaries and/or self-rated questionnaires are widely used to analyze the individual's subjective SQ; the PSQI \cite{buysse1989pittsburgh} is a typical tool used in this direction. Sleep related measures are collected using the individual reported questionnaires based on the subject's perception over a long-term interval \cite{chang2020isleep}. A single-item SQ scale (SQS) was newly developed in \cite{snyder2018new} and provided a simple and practical alternative for the subjective evaluation.
Despite the potential bias, this subjective method is easy to be implemented and has shown useful in various studies \cite{mendoncca2019review}.  

Meanwhile, the prevalence of clinical and commercial IoT devices provide the possibility to measure the SQ objectively. Advanced sensing techniques enable to record various signals of the human body; popular deep learning techniques offer new ideas for data analysis \cite{gan2022knowledgeinteraction,gan2022knowledgestructure}.
These provide more promising solutions to explore SQ in a date-driven manner. PSG is recognized as the gold standard for SQ studies by capturing multiple channels of psychological states, but is limited to the sleep clinics \cite{chang2020isleep}. Actigraphy method using unobtrusive wristbands to collect the body movement data, is an inexpensive alternative, but it measures the SQ only based on the body movement data which limits its usage in the real applications \cite{upadhyay2020personalized}. Hence many researches investigate to use the holistic data (psychological, behavioral and subjective data, etc.) on this task \cite{mendoncca2019review,lim2022assessing,nguyen2021models}. 
Park et al. \cite{park2019learning} proposed a LSTM-Attention method for predicting the SQ using heterogeneous data, including sleep records, daily activities, and demographics.  Bai et al. \cite{bai2012will} developed a prototype system called SleepMiner to collect on-phone data, such as daily activity, living environment and social activity, and used a factor graph model to predict SQ. 
Sathyanarayana et al. \cite{sathyanarayana2016sleep} predicted good or poor sleep efficiency as an indicator of SQ using the physical activity wearable data during awake time. 
Liu et al. \cite{liu2019prediction} predicted SQ based on users' physical exercise data. 
Liu et al. \cite{liu2019gait} used the Kinect sensor to capture the gait pattern to predict SQ, and showed the feasibility of the gait pattern for revealing SQ. 
Krishna et al. \cite{krishna2016sleepsensei} combined features related to user surroundings and movements during sleep, and used regression models to monitor SQ and determine the apt sleep duration. 

Some researchers developed IoT-based sleep monitoring system or mobile apps to monitor the SQ. SleepApp \cite{ravichandran2020sleepapp}, SleepCoacher \cite{daskalova2016sleepcoacher}, SleepTight \cite{choe2015sleeptight} and iSleep \cite{chang2020isleep} are typical mobile-based self-experimentation systems for SQ. Reference \cite{saleem2020iot,milici2018wireless,peng2007multimodality,kim2020iot,zhang2019smars,lin2016sleepsense,laurino2020smart} proposed various IoT based hardware and software solutions to monitor SQ based on multimodal sensors.   

Different from the above researches, this paper monitors the individual SQ based on both the objective and subjective data from multiple sources, and move a step further to provide personalized feedback to improve the SQ in a data-driven manner.




\begin{figure*}[!tbp]
\centerline{\includegraphics[width=\textwidth]{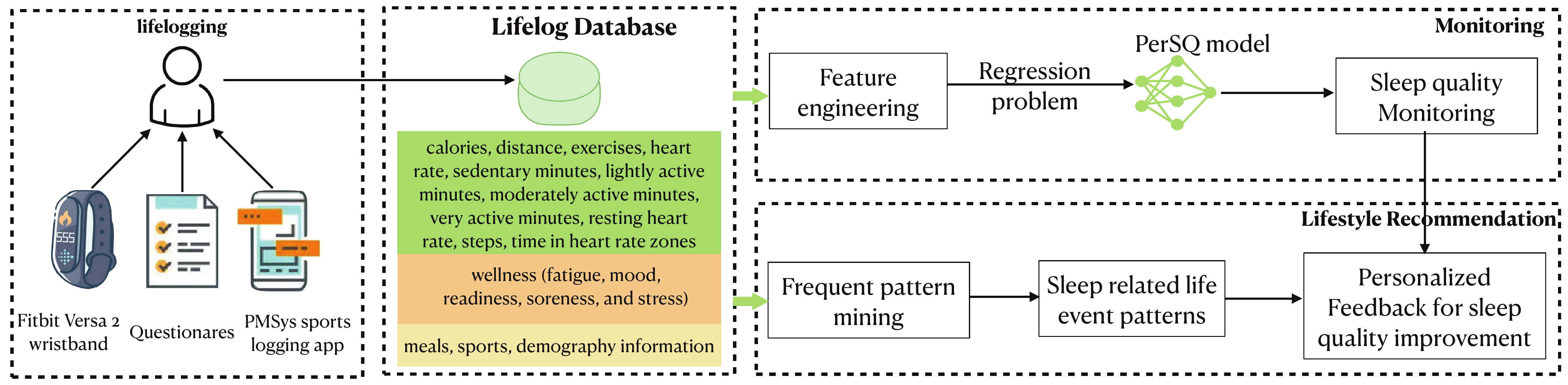}}
\caption{Our framework for monitoring and improving SQ from life-logs.}
\label{Fig:framework}
\end{figure*}

\subsection{Life Event and SQ}

Various studies have explored the impact of various life events on sleep, for example the physical activities and the sleep context \cite{sathyanarayana2016impact,park2019learning,pandey2020personalized,upadhyay2020personalized}. Pandey et al. \cite{pandey2020personalized} 
and Upadhyay et al. \cite{upadhyay2020personalized} conducted N-of-1 experiments to identify the casual relationships between the daytime activities and SQ.  
Park et al. \cite{park2019learning} modeled SQ from daily activities and demographics. 
Singh et al. \cite{singh2021paris} explored the relationship between physical activity and SQ and extracted the behavior modes that resulting in the good sleep using clustering technique. 
Liu et al. \cite{liu2019prediction} predict the SQ only based on the users' daily exercise information. 
Ravichandran et al. \cite{ravichandran2020sleepapp} explored the evidence of individual behaviors affect the SQ, and also showed how sleep affect daytime cognitive function (daytime alertness and working memory). Farajtaba et al. \cite{farajtabar2019modeling} conducted a large-scale many-to-many factor study of  sleep effects and exercise, search query logs, location and aggregated social media data. All these researches show the close relations between the SQ and various life events, providing the references for this paper to choose variables in SQ monitoring. 

Compared with these existing researches, this paper builds the connections between
SQ and the daily activities, the subjective wellness and personal demographic data, with a focus on the timely SQ monitoring before
bed time and providing feedback to improve it. Hence it intentionally avoids using the sleep related data during the current day, an unique feature to enable the real-time and make it different from existing studies. 


\section{Our Proposed Framework}

This paper proposes a novel framework for the SQ monitoring and improvement from long-term lifelogs (Figure \ref{Fig:framework}). 
The lifelogs are collected from three aspects: the daily activity and psychological data from the Fitbit wristband, the individual wellness data logged through PMSys app, and the daily reporting data (meals, body weight, etc.) by everyday questionnaire. Using these lifelogs, our framework conducts the SQ monitoring and improvement in a consecutive manner. After conducting feature engineering on the original data, we model the SQ monitoring as a regression problem and propose a deep learning based personal SQ model, named PerSQ, to monitor the SQ from long-term personal heterogeneous data. For improving the personal SQ, we conduct frequent pattern mining on the whole lifelogs and generate three sets of sleep patterns (i.e., low, normal and high ones). Based on the predicted personal SQ, we match the user data with the obtained sleep patterns to find the parameters to be optimized and provide feedback to the individuals for further SQ improvement.

\subsection{Personalized Long-term SQ Monitoring}

\noindent\textbf{Problem Formulation}
For an individual user $u_{i}$, this research aims to predict his/her SQ on the current day $t_{m}$ based on the objective data (i.e., daily activity $da$, personal demographic information $pd$) and subjective data (i.e., wellness $ws$) collected from multiple sources in a long term. Following existing researches \cite{crivello2019meaning}, we use the sleep efficiency as the index of SQ calculated as follows:
\begin{equation}
    SQ = \frac{minutes\_Asleep}{time\_in\_bed} \in [1,100]
\end{equation}
Let $X_{m} =\{da_{m}, pd, ws_{m}\}$ be the list of all data obtained for a day $t_{m}$, and $SQ_{m}$ be the value of SQ on the same day. The problem of SQ monitoring is formulated as:
\begin{equation}
    f(X_{m-t},X_{m-t+1},...,X_{m-1},X_{m}) \longrightarrow SQ_{m}
\end{equation}
where $t$ is the number of previous days considered, and $f$ is the  function to predict the SQ value. We consider the carry-over influence of previous days on the current day.  

\begin{figure}[!tbp]
\centerline{\includegraphics[width=.5\textwidth]{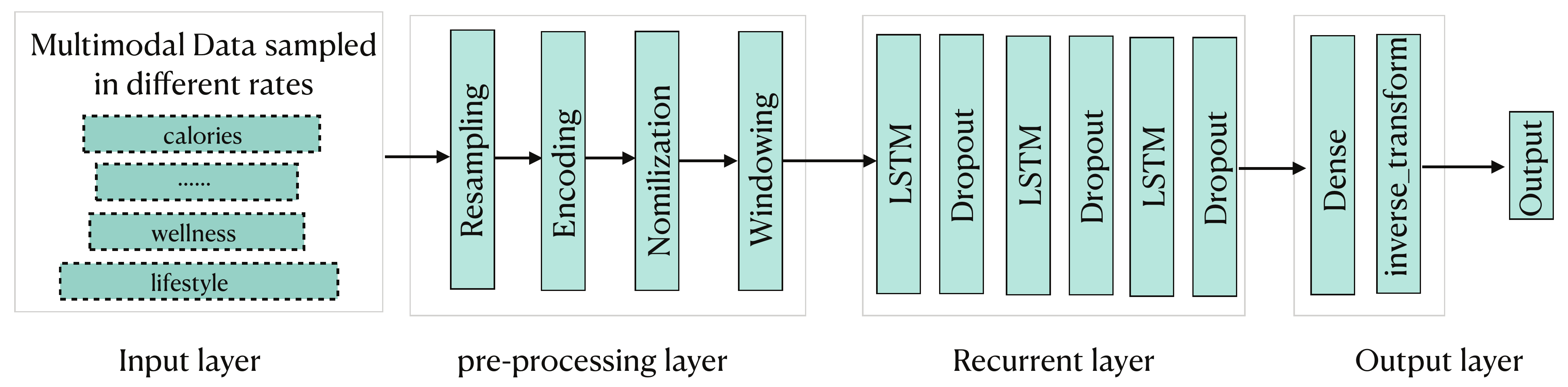}}
\caption{The architecture of the proposed PerSQ model.}
\label{Fig:PerSQ_model}
\end{figure}

\noindent\textbf{PerSQ Model} The architecture of the proposed PerSQ model is shown in Figure \ref{Fig:PerSQ_model}. It mainly consists of three layers: the pre-processing layer, the recurrent layer and the output layer. 

\textit{Pre-processing layer:} we re-sample the input data into the day interval and encode the category variables using one-hot encoding. Considering the variables are in different units, we normalize the data into [0,1] using min-max normalization. To capture the carry-over effects, we shift the long-term data using a windowing mechanism, and slide the fix-length window to generate the time-series data $X = <X_{m-t},X_{m-t+1},...,X_{m-1},X_{m}>$.

\textit{Recurrent layer:} Long short-term memory (LSTM) network is used in this layer to extract useful information from the sequences. We uses three LSTM networks with dropout to process the time-series data. 

\textit{Output layer:} Given the processed data $v$ from the recurrent layer, we firstly map $v$ into the target space using fully-connected layer to obtain $v^{'}$. To map the final predicted value $v^{'}$ to the original scale (i.e., [1,100]), we concatenate it with the original data $X$ and use inverse transformation to make it in the original scale. Hence the final output for the predicted SQ is in range of 1 and 100. 
\begin{equation}
\begin{aligned}
    &v^{'} = Wv+b\\
    &\hat{SQ} = inverse\_transform(X, v^{'})\in [1,100]
\end{aligned}
\end{equation}

\subsection{Life Event Pattern Mining}

To obtain the SQ insights from daily life events, we conduct pattern mining on the lifelogs to discover the different sets of patterns. Our hypothesis is that these patters contain the relations of SQ and daily life events and can be used as references to provide further lifestyle feedback. 

\begin{figure}[!tbp]
\centerline{\includegraphics[width=.3\textwidth]{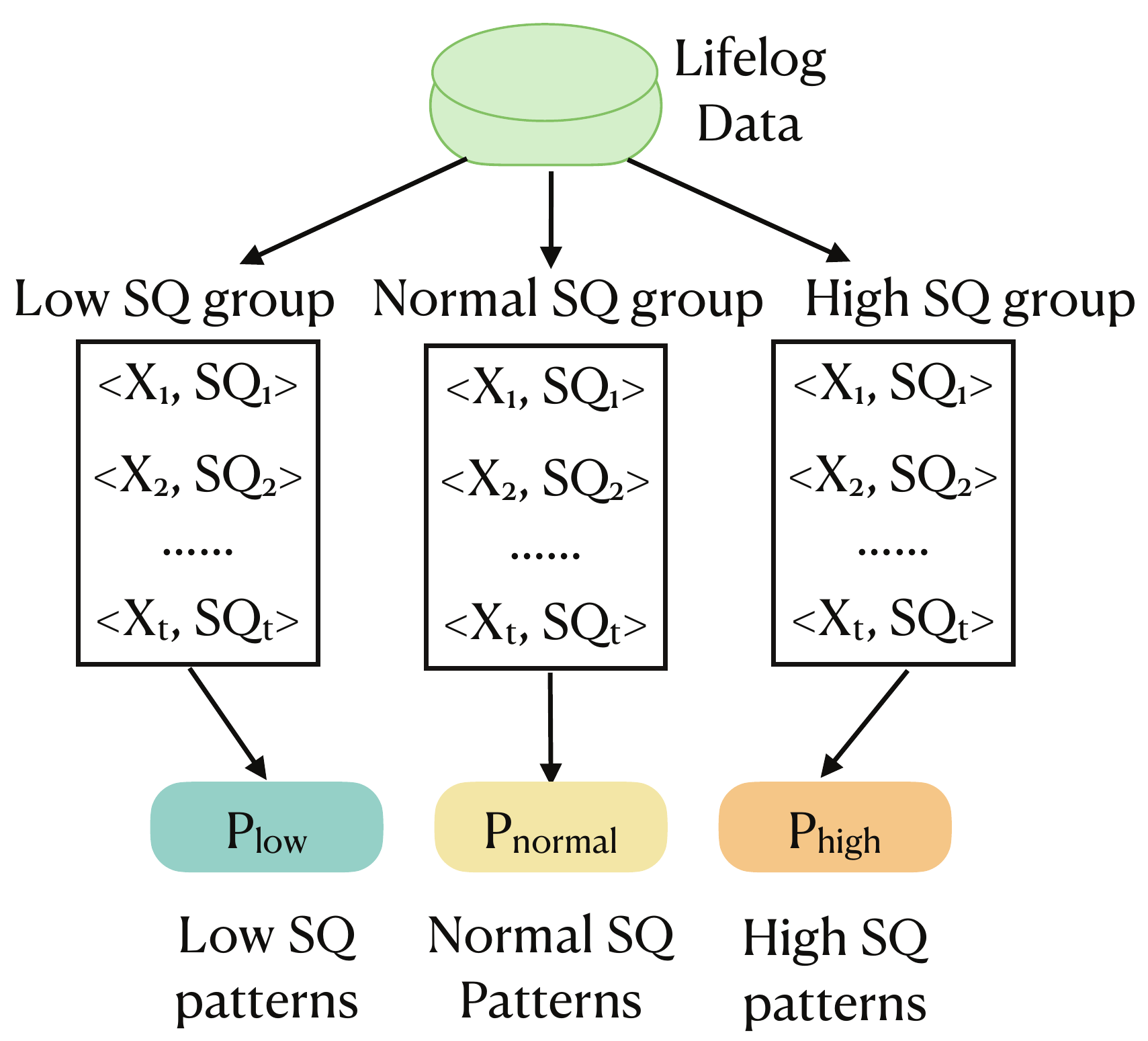}}
\caption{Lifelog mining for obtaining three sets of SQ patterns.}
\label{Fig:mining}
\end{figure}

To this aim, we divide the lifelogs of all persons into three groups: the high, normal and low SQ groups based on the distribution of SQ values, as shown in Figure \ref{Fig:mining}. The pattern mining process is then conducted on each group to find the frequent patterns in different degree. Specifically, we first select a group of variables, which are related with the daily activities and individual wellness, and infer the potential patterns with daily SQ. The group of variables for our dataset is shown in Figure \ref{Fig:thresholds}. These life event factors are then thresholded to transform them from continuous to the category ones for easier working with the pattern mining method.  
The Apriori algorithm \cite{agrawal1994fast} is adopted to mine the frequent association rules from these three sets of items. Finally, three groups of frequent patterns are obtained, providing the insights for understanding the impact of daily life events on different degrees of SQ. The results of pattern mining are shown in Section \ref{section:Frequent_Patterns}.

\subsection{Personalized Lifestyle Feedback for SQ Improvement}

After obtaining a user's daily data, we can predict his/her SQ using the proposed PerSQ model. Based on the predicted value (low, normal, high), personal lifestyle feedback is conducted using the life event patterns for improving their SQ. 

Inspire by \cite{pandey2020personalized} to use condition-action rules for health recommendations, we use the obtained frequent life event patterns to generate the personal feedback. Since these event patterns contain the daily life context information linking to different degrees of SQ, they provide the optimization insights for improving SQ.  
We abstract and match the personal data with the event patterns using a simple rule matching strategy based on the max co-occurrence of the same contextual parameters. Meanwhile we rank the matched rules with the support count and the length of the variables in the dataset to enable the better matching results. For the low predicted SQ, we match the user data with the event patterns in the normal and high group; while for the normal and high results, we match the data with the patterns only in the high group, with the aim of potential improving the personal SQ based on the superior patterns. After obtaining the proper matched pattern, we discover the contextual parameters from the unmatched parts in the pattern and generate the feedback based on these optimizable parameters. For reference, the algorithm for this feedback process is shown in Alg. \ref{Feedback_algorithm}.
Finally, the actionable multiple-item feedback (including the predicted SQ score and the suggestions) can be presented to the users for improving their personal SQ.

\begin{algorithm}[tb]\small
\caption{Personalized Lifestyle Feedback}
\label{Feedback_algorithm}
\begin{algorithmic}[1]
\Require A user's time-series data $X = <X_{m-t},X_{m-t+1},...,X_{m-1},X_{m}>$, the trained SQ monitoring model $PerSQ$, the mined three sets of frequent patterns $P_{low},P_{normal},P_{high}$. 
\Ensure The predicted SQ value $SQ_{m}$ and the feedback $f$. 
\State initialize the set $\mathbb{P}=\emptyset$ for parameters to be optimized;\\
\textbf{Sleep quality Prediction:}
\State $SQ_{m} \longleftarrow PerSQ(X)$;\\
\textbf{Personalized Feedback:}
\State threshold the personal data to category ones: $X^{'}\longleftarrow X$;
\If{$SQ_{m}$ is in low SQ group}
    \State  $\mathbb{P} \longleftarrow$ $RuleMatch(X^{'}, P_{normal} \& P_{high})$;
\ElsIf{$N$ is normal and high SQ group}
    \State  $\mathbb{P} \longleftarrow$ $RuleMatch(X^{'}, P_{high})$;
\EndIf 
\State  generate $f$ based on $\mathbb{P}$;
\State return $SQ_{m}$ and $f$.
\end{algorithmic}
\end{algorithm}

\begin{table*}[t]\tiny
\centering
\caption{The multimodal data sources used for the study.}
\label{transcription_protocol}
\tiny
\resizebox{1\textwidth}{!}{ %
\begin{tabular}{llc}
\hline
\textbf{Data type (Multimodalities)}                                   & \textbf{Description}                                                                                                                                                     &  \textbf{Mean ±SD} \\ \hline 
Daily Activity (Source: Fitbit)             &                                                                                                                                                                                       &          \\
\qquad calories                                    & calories the person has burned per day                                                                                                                 &  3053.2±742 .3       \\
\qquad distance                                    & the distance (in centimeters) moved per day                                                                                                                  &  950694.3±505781.7         \\
\qquad steps                                       & the number of steps per day                                                                                                                                  &  12221.6 ± 5959.0       \\
\qquad sedentary minutes                          & the number of sedentary minutes per day                                                                                               &    685.1±172.2      \\
\qquad lightly active minutes                    & the number of lightly active minutes per day                                                                                                                                &   229.4±95.7        \\
\qquad moderately active minutes                 & the number of moderately active minutes per day                                                                                                                 &   24.4±23.0        \\
\qquad very active minutes                       & the number of very active minutes per day                                                                                                                       &   49.0 ±44.4      \\
\qquad time in heart rate zones                & the number of minutes in different heart rate zones (4 types)                                                                                                             &           \\ \hline
wellness (Source: PMSys App)                &                                                                                                                                                                                      &          \\
\qquad fatigue                                     & 1-5 scale of the subjective measurement of fatigue                                                                                                                              &  2.7±0.7        \\
\qquad mood                                        & 1-5 scale of the subjective measurement of mood                                                                                                                               &   3.2±0.7        \\
\qquad readiness                                   & 1-10 scale of how ready to exercise                                                                                                                                           &    4.9±1.9       \\
\qquad soreness                                    & 1-5 scale of the subjective measurement of scoreness                                                                                                                          &   2.8±0.6        \\
\qquad stress                                      & 1-5 scale of the subjective measurement of stress                                                                                                                              &  2.9±0.7         \\ \hline
Personal Demographic (Source: Questionaire) &                                                                                                                                                                                     &          \\
\qquad age                                         & the age of a user                                                                                                                                               &   35.6±11.4        \\
\qquad gender                                      & the gender of a user                                                                                                                                                                    &   male:12, female:3       \\
\qquad A or B                                      & \begin{tabular}[c]{@{}l@{}}whether the person has a Type A or Type B personality;\\ getup early in the morning (Type A) and wakes up late (Type B)\end{tabular}                  &   A:9, B:6       \\ \hline
sleep informaition (Source: Fitbit)         &                                                                                                                                                                                      &          \\
\qquad minutes Asleep                              & the number of minutes being asleep                                                                                                                                       &          \\
\qquad time in bed                                 & the number of minutes in bed                                                                                                                                    &           \\ 
\qquad sleep quality                                 & the quality of sleep calculated by minutes Asleep / time in bed                                                                                                                                                    & 87.8± 3.8 \\       \hline
\end{tabular}}
\end{table*}

\begin{figure*}[!tbp]
\centerline{\includegraphics[width=\textwidth]{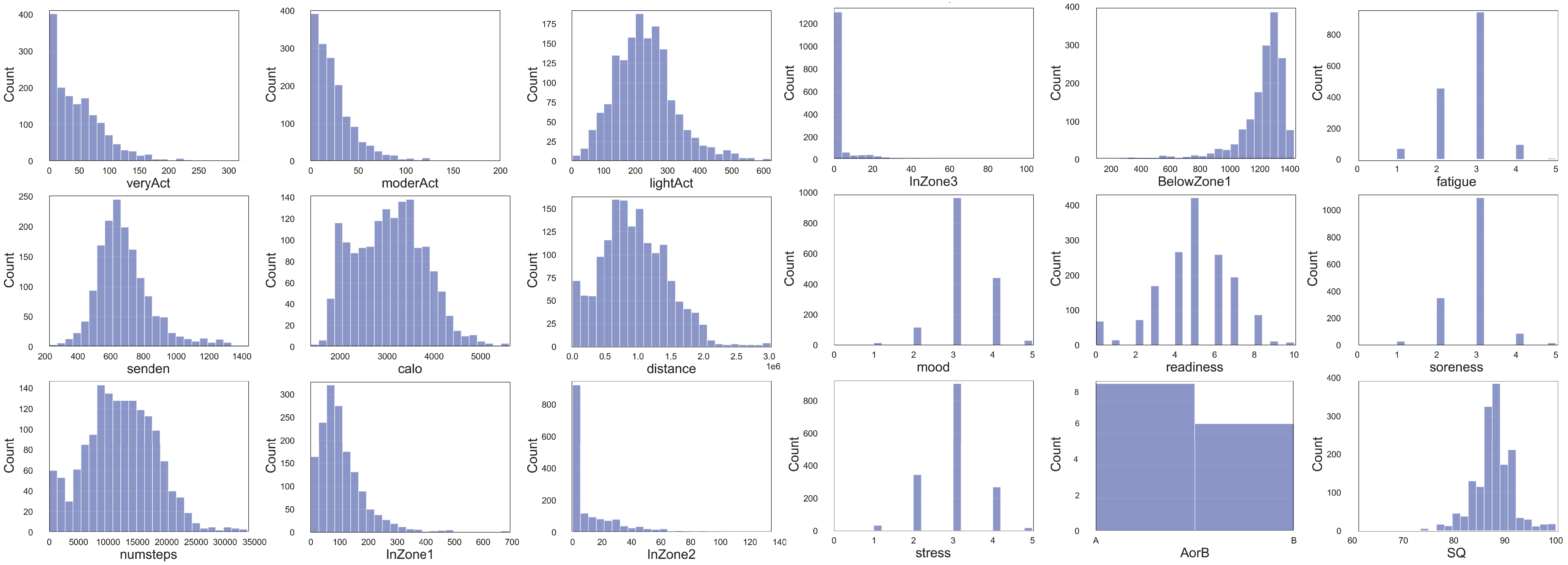}}
\caption{The distributions of different life event variables.}
\label{Fig:distributions}
\end{figure*}

\section{Dataset and Preprocessing}

\subsection{Dataset}

The PMData\footnote{PMData: \url{https://datasets.simula.no/pmdata/}}\cite{thambawita2020pmdata} is used for this study. 
It is a lifelogging dataset combining the traditional lifelogging with sports activity logging from 16 persons during 5 months (November 2019 to March 2020) using Fitbit Versa 2 smartwatch wristband, the PMSys sports logging app, and Google forms for the multimodal data collection. 
The objective physical and physiological data, such as the calories, steps, distance, exercise, heart rate, different kinds of activity minutes,  sleep information and heart rate measurements, are recorded using Fitbit Versa 2 smartwatch.  
Personal wellness, including parameters like fatigue, mood, readiness, soreness and stress, is subjectively measured and logged by users through the PMSys sports logging app. Sports, meal and personal information are reported daily using the Google forms. Eventually, there are 2,440 activity sessions, 20,991,392 heart rate measurements, 1,836 days of sleep scores and 1,747 wellness reports in the final dataset.

As the 17 overall types of logged data have different rate of entries (per minutes, per day, per 5 seconds, etc.), we unify them into the same day interval. After checking the missing values, we find one user (ID:p12) did not have the record about light active minutes, we remove the data from this user and use the other 15 participant data in the final experiments. We select the factors that are highly related with the SQ task in previous studies, and finally get 8 factors belong to daily activity, 5 factors for the personal wellness, and 3 factors for the personal demographic (as shown in Table \ref{transcription_protocol}). The sleep information collected using the Fitbit are used as the ground-truth to build our model. 

The distributions of these factors are shown in Figure \ref{Fig:distributions}. Among the 15 persons, 12 are male, the average age is 35.6 years old. Nine of them have the habit of getup early in the morning. The factors in the category of daily activity are transformed into the day interval with different distributions. For the factors in wellness, fatigue, mood, soreness and stress use the 5-point Likert scale to measure the subjective feeling of participants, while the readiness uses 10 scale. The ground-truth SQ values for all the people have an average of 87.8.

\begin{figure*}[!tbp]
\centerline{\includegraphics[width=\textwidth]{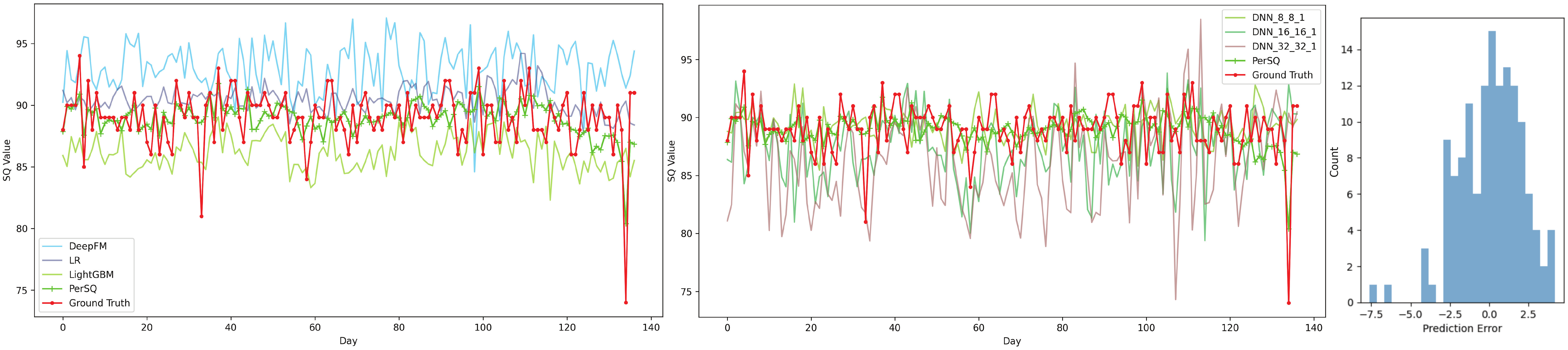}}
\caption{(left): Comparisons of different models on SQ prediction for an individual user; (right): the distributions of prediction errors of the PerSQ model.}
\label{Fig:Comparisons}
\end{figure*}

\section{Experiments}

\subsection{Experimental Setup}

We trained the PerSQ model for SQ modeling using leave-one-out cross validation on the preprocessed dataset. Grid-search over heuristic choices of hyper-parameters was performed over choices of the numbers of  sub-layers of LSTM networks and the number of days to track back. Eventually the cell units for the three consecutive sub-layers of LSTM in our model were set as $<50,30,20>$ and the model performances were compared on different considered days. For a new user who has no long‐term event logs, we initialized the personalized model using the trained model at the first stage, and then updated it when more personalized data was collected to make it more tailored to the individual user.

To verify the effectiveness of the proposed PerSQ model on SQ monitoring, we compared with several competing models that have been shown to be effective in previous researches \cite{mendoncca2019review,lim2022assessing}: the linear regression (LR) \cite{ravichandran2020sleepapp}, the LightGBM \cite{liu2019prediction}, the multilayer perceptron (MLP) network \cite{sathyanarayana2016impact} and the deep factorization machine (DeepFM) \cite{guo2017deepfm}. To fairly compare these models with the proposed model, we also take consideration of the carry-over effect by adding the SQ value in the previous day as a variable to input into these models.  
All the models were implemented in Python, the optimal hyperparameters were tuned to report the best results. Four criteria are employed to evaluate different facets of the regression performance: the Mean Absolute Error (MAE), Mean Squared Error (MSE), Root Mean Square Error (RMSE) and the $R^2$. The first three tell how well a regression model can predict the value of SQ while the $R^2$ is a statistical measure that tells how well the model fit on the original data.
Moreover, to validate the lifestyle feedback, we performed case study to show how to leverage the generated frequent patterns to improve the individual SQ.

\subsection{Results on SQ Monitoring}


\begin{table}[!t]
\centering
\caption{Performance comparison of different models.}
\label{tab:Performance_comparison}
\tiny
\resizebox{0.5\textwidth}{!}{ %
\begin{tabular}{ccccc}
\hline
Model                                       & MAE   & MSE    & RMSE  & $R^2$  \\ \hline
Linear regression                           & 2.040 & 8.148  & 2.856 & -0.511 \\
LightGBM                                    & 3.217 & 13.871 & 3.724 & -1.571 \\
\multicolumn{1}{l}{MLP regression(32-32-1)} & 4.035 & 26.804 & 5.177 & -3.969 \\
MLP regression(16-16-1)                     & 3.032 & 15.551 & 3.944 & -1.883 \\
MLP regression(8-8-1)                       & 1.869 & 7.033  & 2.652 & -0.304 \\
DeepFM                                      & 4.164 & 24.874 & 4.987 & -3.611 \\
\textbf{PerSQ}                                       & \textbf{1.591} & \textbf{4.068} & \textbf{2.017} & \textbf{0.246}  \\ \hline
\end{tabular}}
\end{table}

\begin{figure}[!tbp]
\centerline{\includegraphics[width=.3\textwidth]{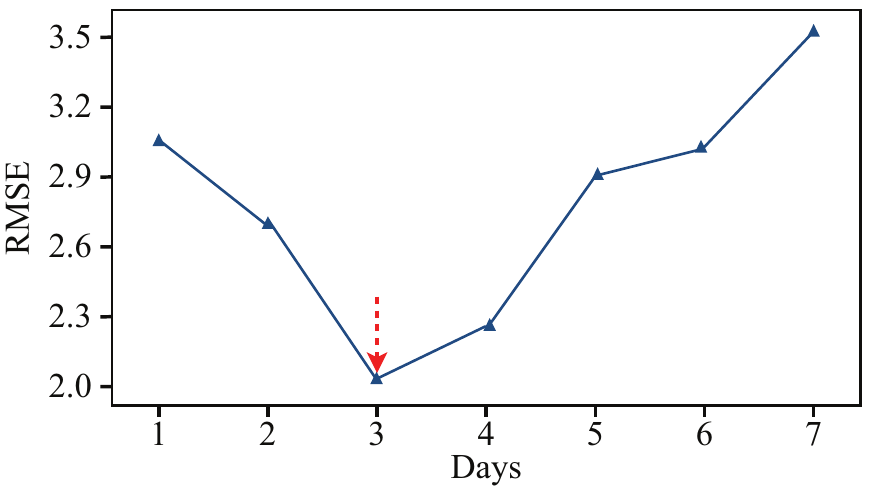}}
\caption{The model performances considering different numbers of days.}
\label{Fig:days_considered}
\end{figure}

\begin{table*}[!t]
\centering
\caption{Some examples of the life event patterns obtained for three groups.}
\label{tab:event_patterns}
\tiny
\resizebox{\textwidth}{!}{ %
\begin{tabular}{clccc}\hline
\multicolumn{1}{l}{SQ level} & \multicolumn{1}{c}{Some examples of life event patterns} & Frequency & Total number & Insights \\ \hline
\multirow{6}{*}{\begin{tabular}[c]{@{}c@{}}Low\\ (251)\end{tabular}}    & ('soreness\_normal', 'InZone3\_normal', 'numsteps\_low', 'distance\_low', 'veryAct\_low', 'InZone2\_low')  & 48 &\multirow{6}{*}{46}&\multirow{6}{*}{\begin{tabular}[c]{@{}c@{}}Physical\\ Activities\end{tabular}}\\
& ('numsteps\_low', 'InZone3\_normal', 'moderAct\_low', 'distance\_low', 'veryAct\_low', 'InZone2\_low')  &  47&&\\
& ('InZone1\_low', 'InZone3\_normal', 'numsteps\_low', 'distance\_low', 'veryAct\_low', 'InZone2\_low')  &  46&& \\
& ('InZone3\_normal', 'numsteps\_low', 'distance\_low', 'veryAct\_low', 'InZone2\_low')  &  58 &&\\
& ('soreness\_normal', 'InZone3\_normal', 'numsteps\_low', 'distance\_low', 'InZone2\_low')  & 55 &&\\
& ('readiness\_normal', 'InZone3\_normal', 'numsteps\_low', 'distance\_low', 'InZone2\_low') & 51 &&\\\hline
\multirow{6}{*}{\begin{tabular}[c]{@{}c@{}}Normal\\ (998)\end{tabular}} &('soreness\_normal', 'mood\_normal', 'stress\_normal', 'fatigue\_normal') &  272&\multirow{6}{*}{52}&\multirow{6}{*}{Wellness}\\ 
&('soreness\_normal', 'mood\_normal', 'InZone3\_normal', 'stress\_normal')  &  267 &&\\
&('mood\_normal', 'InZone3\_normal', 'stress\_normal', 'AorB\_A')  &  264&& \\
&('soreness\_normal', 'InZone3\_normal', 'stress\_normal', 'fatigue\_normal')  &  256& &\\
&('soreness\_normal', 'mood\_normal', 'stress\_normal')  &  340 &&\\
&('soreness\_normal', 'mood\_normal', 'fatigue\_normal') &  339 &&\\\hline
\multirow{6}{*}{\begin{tabular}[c]{@{}c@{}}High\\ (322)\end{tabular}}&('soreness\_normal', 'numsteps\_normal', 'InZone3\_normal', 'distance\_normal', 'veryAct\_normal')  & 86 &\multirow{6}{*}{49}&\multirow{6}{*}{\begin{tabular}[c]{@{}c@{}}Physical\\ Activities,\\lifestyle,\\ wellness\end{tabular}}\\
&('soreness\_normal', 'numsteps\_normal', 'InZone3\_normal', 'fatigue\_normal', 'distance\_normal') &  82& &\\
&('AorB\_A', 'numsteps\_normal', 'InZone3\_normal', 'distance\_normal', 'veryAct\_normal')  & 76&& \\
&('AorB\_A', 'soreness\_normal', 'numsteps\_normal', 'InZone3\_normal', 'distance\_normal')  &  75&& \\
&('InZone3\_normal', 'veryAct\_normal', 'numsteps\_normal', 'distance\_normal') &  109&& \\
&('InZone3\_normal', 'stress\_normal', 'numsteps\_normal', 'distance\_normal')  & 100&&\\ \hline
\end{tabular}}
\end{table*}

The different models were evaluated by their performances in monitoring the SQ on the PMData. As show in Table \ref{tab:Performance_comparison}, the proposed PerSQ model achieves the best performance on all the four criteria, with RMSE score of 2.017 and $R^2$ score of 0.246. The MLP regression model with 8-8-1 network architecture obtains the second best, with RMSE score of 2.652. The 
variations of MLP with 32-32-1 and 16-16-1 architectures do not get better results with deeper neurons, this is probably because the dataset is not big enough to train these models well. Even with the linear regression model, it performs much better than the two variations. LightGBM and DeepFM, two of the powerful models that rank in the top of various Kaggle competitions, seems do not get better performance than the linear regression model on this task. By contrast, the proposed PerSQ model takes consideration of the dependencies of previous life events on the SQ of the current day, making it perform better for SQ monitoring in a long term. 

Moreover, we compared the performance of these different models for monitoring the individual SQ. As shown in Figure \ref{Fig:Comparisons}(left), the predicted SQ scores of an individual user by different models are visualized in different colors. The red line is the ground-truth, the prediction of our PerSQ model is shown in dark green. Compared with other models, the PerSQ is more closer to the ground-truth. The prediction error of the PerSQ is shown in Figure \ref{Fig:Comparisons}(right); most of the predicted errors are in range [-2.5, 2.5], making it acceptable for real applications.

To test the influence of life events in previous days, we tested the model performance using data from different numbers of days. As shown in Figure \ref{Fig:days_considered}, the RMSE scores of different considered days are compared. PerSQ model archives best performance when considering three previous days; the RMSE score decreases when considering more days, and afterwards increases gradually when more than three days are considered.

\begin{figure}[!tbp]
\centerline{\includegraphics[width=.5\textwidth]{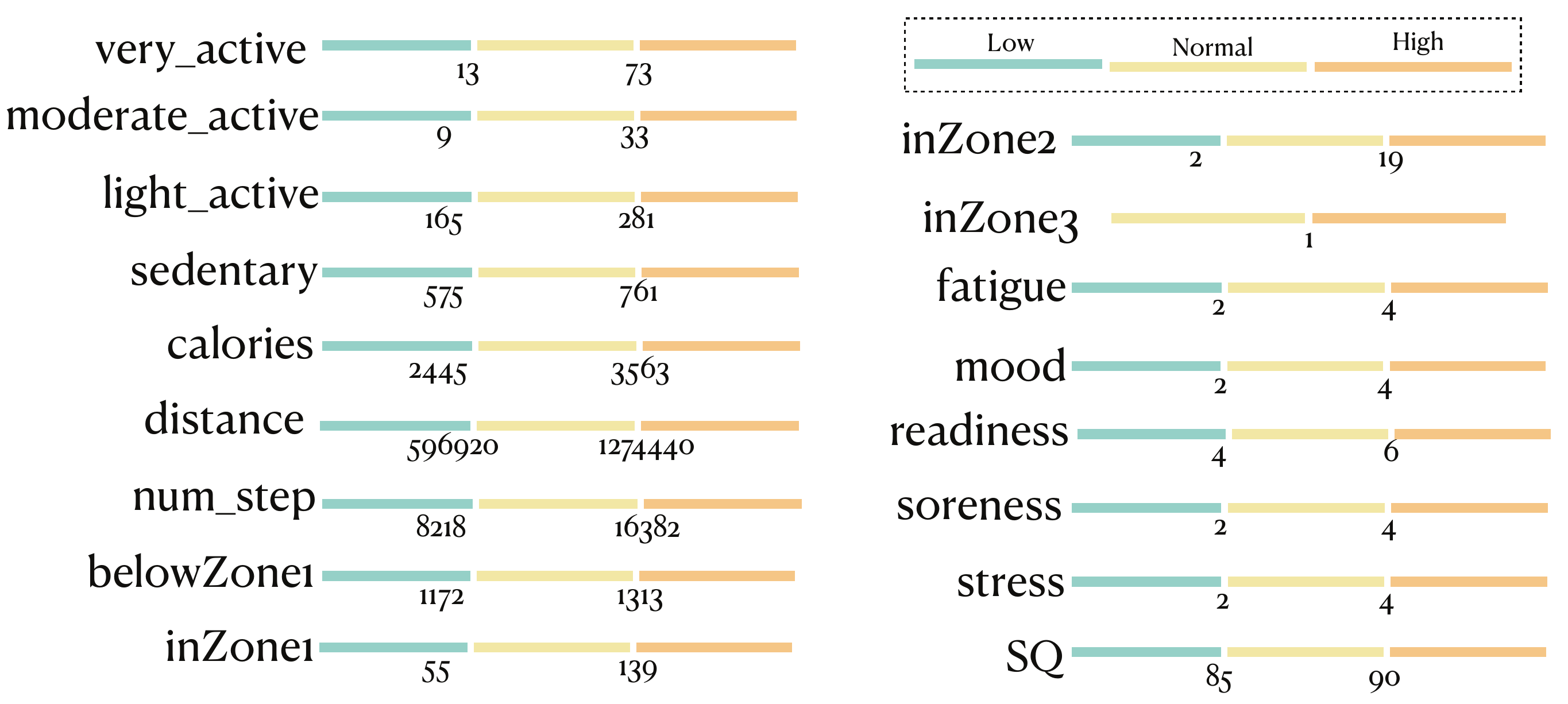}}
\caption{The thresholds for different life event variables.}
\label{Fig:thresholds}
\end{figure}

\begin{table*}[!t]\tiny
\centering
\caption{Feedback designing to optimize life parameters for improving SQ.}
\label{tab:feedback_design}
\resizebox{\textwidth}{!}{ 
\begin{tabular}{cll}
\hline
Category                                                                        & Parameter to optimize* & Feedback examples \\ \hline
\multirow{6}{*}{\begin{tabular}[c]{@{}l@{}}Physical \\ activities\end{tabular}} & numsteps          &   ``Please try to walk more''       \\
 &    distance                    &  ``Let's go out and have a walk''        \\
&   calories                    &  ``Let's do something to consume more calories''        \\
 &   very\_active or inZone3                  &   ``Recommended activities: Running, jogging, race walking/Hiking/Fast biking/ Stair climbing, etc. ''       \\
 &   moderate\_active or inZone2                  &  ``Recommended activities: Brisk walking/Ball sports/Water aerobics/Ballroom dancing, etc.''        \\
 &   light\_active or inZone1                   & ``Recommended activities: Slow walking/Slow bike riding/Stretching exercise/Gentle or chair yoga, etc.''         \\
\hline
\multirow{5}{*}{wellness}                  &  fatigue                     &  ``You seems tired, let's take some rest''        \\
&  mood                     &  ``Let's adjust our mood, how about listening to a happy music?''        \\
&  readiness                     &  ``How about taking a deep breath and making ourselves ready?''        \\
&  soreness                     &  ``You may need some relex''        \\
&    stress                   &     ``Go easy on yourself, let's make time for hobbies''      \\ \hline
lifestyle                                  &  AorB: B$\rightarrow$A                     &   ``Do you want to try to keep early hours?''       \\ \hline
\multicolumn{3}{l}{*The optimization is to improve the level of parameters: low$\rightarrow$normal$\rightarrow$high.} 
\end{tabular}
}
\end{table*}

\subsection{Frequent Patterns of SQ and Life Events}\label{section:Frequent_Patterns}

To discover the insights of life event patterns with daily SQ, we threshold the considered factors to transform them from continuous to the category ones, as shown in Figure \ref{Fig:thresholds}, based on their own distributions in Figure \ref{Fig:distributions}.

Based on the thresholded SQ value, we divided all the samples into three group: 251 samples in the low level, 998 in the normal level and 322 in the high group. After conducting the Apriori algorithm on each group, we obtained the frequent life event patterns in each group. Table \ref{tab:event_patterns} shows some examples of these patterns with highest frequencies. We filtered the patterns with at least 20\% percent of the total number as the minimum support count. After pruning the frequent itemsets, we retained the longest and second-longest patterns and used them for future lifestyle feedback. 
Finally, we got 46, 52 and 49 life event patterns for these three SQ groups. 

The differences of these frequent patterns in three groups present some insights for exploring different levels of SQ.  
For the low group, most of the patterns are with low values of variables, and most of these variables are related with the indicators of physical activities, such as the heart beat zones, the number of steps, the distance and the very and moderate active minutes. These patterns generally show the low SQ is greatly related with low level of physical activities. For the normal SQ group, most of the variables in the patterns are related with the human wellness, such as the soreness, mood, stress and fatigue. While for the high SQ group, the variables are not only related with the physical activities but also the lifestyle (i.e., keep early hours) and wellness factors. This is consistent with existing researches that regular exercise, proper lifestyle and good wellness could help people get a high quality of sleep.

\subsection{Case Study}

To test the effectiveness of the life event patterns for providing SQ improvement feedback, we designed the feedback items in various contexts and conducted a case study to show how they works. 

\begin{figure}[!tbp]
\centerline{\includegraphics[width=0.5\textwidth]{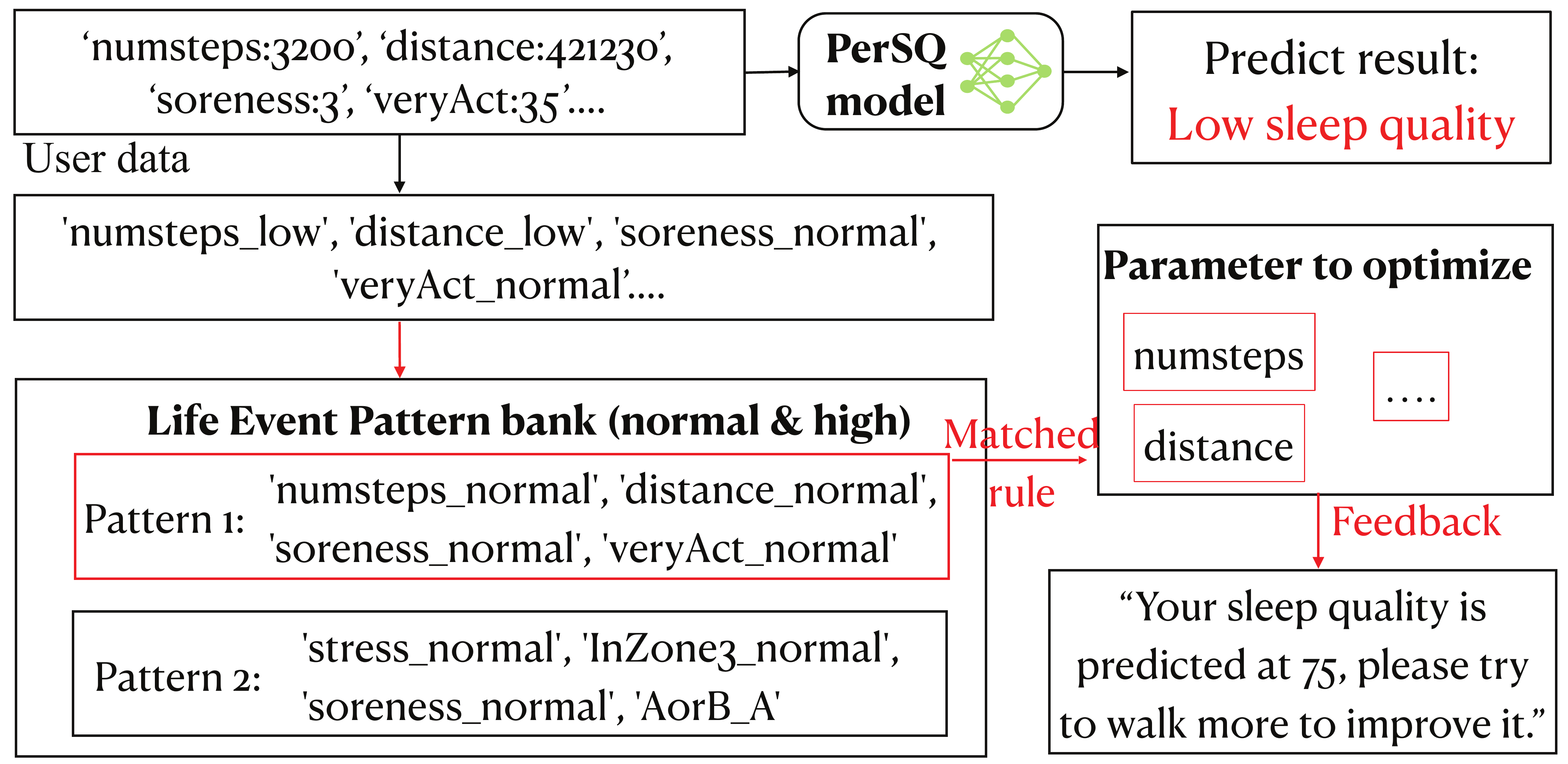}}
\caption{Case study of personalized feedback for a user.}
\label{Fig:feedback}
\end{figure}

As shown in Table \ref{tab:feedback_design}, we designed some example feedback in three categories based on the report in  \cite{USHHS2018Physical} to optimize the life parameters to improve the individual SQ. The feedback can be provided to users as voice or messages through the mobiles or wristwatches. 
Based on the designed feedback and the discovered life event patterns, we conducted a case study using the data from a user, as illustrated in Figure \ref{Fig:feedback}. 

The piece of data is input to our model and predicted as low SQ (SQ score: 75). After thresholding the data and matching it with the life event patterns in normal and high group, we obtain the most proper pattern (pattern 1). By comparing the differences of the user data and the matched pattern, we find the parameter to optimize, i.e., the \textit{numbersteps} and \textit{distance}. The final feedback is then generated by providing the predicted SQ value and also the designed feedback for these two factors referencing Table \ref{tab:feedback_design}. An individual can know not only how his/her SQ will be on tonight but also how to further improve it based on the feedback.

\section{Conclusion}

This paper presented a framework for monitoring and improving SQ from daily life-logs. It monitored the SQ using  both the objective and subjective data from multiple sources (the Fitbit wristband, the sports logging app, and questionnaires), and move a step further to explore the insights from three groups (low, normal, and high SQ) of data to discover the potential patterns between life events and different levels of SQ. Based on these patterns, it further provided personalized feedback to the individuals to improve their SQ in a data-driven manner.
To monitor SQ, it proposed the deep learning based PerSQ model using the long-term personal heterogeneous data and taking consideration of the carry-over effect of previous life events on the SQ of current day. The frequent pattern mining was conducted on the log data to discover the insights of patterns. Finally the user data was matched with these patterns to find the parameters to be optimized and provide feedback to the individuals for further SQ improvement. 

We took the PMData dataset to build and test our framework. The results showed that the proposed PerSQ model achieved the best performance for the SQ monitoring compared with the other baseline models and the best performance was obtained when considering life events in the previous three days. The insights on the patterns showed that low SQ is more likely related with the insufficient physical activities, normal SQ is more linked with personal wellness, while high SQ may requires many factors.   
Moreover, a case study was conducted to show the process of  personalized feedback, providing information for the individuals to improve their SQ in the future.

For future work, we plan to conduct real application studies to validate the effectiveness of the feedback, and further improve the framework in a perception-action dynamic loop.  
Ultimately, we expect that the suggested framework will help people learn about their sleep and further gain better SQ.


%
%
%

\bibliographystyle{unsrt}
\bibliography{mybibliography}

\balance

\end{document}